\documentclass[12pt]{article}

\usepackage{amsmath,amssymb,amsfonts,amsbsy}
\usepackage{cite}
\usepackage{graphicx}
\usepackage{wrapfig}
\def\xb{\overline{x}}

\begin{document}
\addcontentsline{toc}{subsection}{{Cross sections and spin
asymmetries in vector meson leptoproduction.}\\
{\it S.V. Goloskokov}}

%%%%%%% please do not touch these! %%%%%%
\setcounter{section}{0}
\setcounter{subsection}{0}
\setcounter{equation}{0}
\setcounter{figure}{0}
\setcounter{footnote}{0}
\setcounter{table}{0}

\begin{center}
\textbf{Cross sections and spin asymmetries in vector meson
leptoproduction}

\vspace{5mm}

S.V. Goloskokov

\vspace{5mm}

\begin{small}
Bogoliubov Laboratory of Theoretical Physics, Joint Institute for
Nuclear Research, Dubna 141980, Moscow region, Russia
\end{small}
\end{center}

\vspace{0.0mm} % Don't laugh: it does change the spacing!

\begin{abstract}
Light vector meson leptoproduction  is analyzed on the basis of
the generalized parton distributions. Our results on the cross
section and spin effects  are in good agrement with experiment at
HERA, COMPASS and HERMES energies. Predictions for $A_{UT}$
asymmetry for various reactions are presented.
\end{abstract}

\vspace{7.2mm}

In this report, investigation of vector meson leptoproduction is
based on the handbag approach where the leading twist amplitude at
high $Q^2$ factorizes into hard meson electroproduction off
partons and the Generalized Parton Distributions (GPDs)
\cite{fact}. The higher twist (TT) amplitude which is essential in
the description of spin effects exhibits the infrared
singularities, which signals the breakdown of factorization
\cite{mp}. These problems can be solved in our model \cite{gk06}
where subprocesses are calculated within the modified perturbative
approach  in which quark transverse degrees of freedom accompanied
by Sudakov suppressions are considered. The quark transverse
momentum regularizes the end-point singularities in the TT
amplitudes so that it can be calculated.

In the model, the amplitude of the vector meson production off the
proton with positive helicity reads as a convolution of the
partonic subprocess
 ${\cal H}^V$ and GPDs $H^i\,(\widetilde{H}^i)$
\begin{eqnarray}\label{amptt-nf-ji}
  {\cal M}^{Vi}_{\mu'+,\mu +} &=& \frac{e}{2}\, \sum_{a}e_a\,{\cal
  C}_a^{V}\, \sum_{\lambda}
         \int_{xi}^1 d\xb
        {\cal H}^{Vi}_{\mu'\lambda,\mu \lambda}
                                   H_i(\xb,\xi,t) ,
\end{eqnarray}
where  $i$ denotes the gluon and quark contribution, sum over $a$
includes quarks flavor $a$ and $C_a^{V}$ are the corresponding
flavor factors \cite{gk06};
 $\mu$ ($\mu'$) is the helicity of the photon (meson), and $\xb$
 is the momentum fraction of the
parton with helicity $\lambda$. The skewness $\xi$ is related to
Bjorken-$x$ by $\xi\simeq x/2$.  In the region of small $x \leq
0.01$  gluons give the dominant contribution. At larger $x \sim
0.2$  the   quark contribution plays an important role\cite{gk06}.

To estimate  GPDs, we use the double distribution representation
\cite{mus99}
\begin{equation}
  H_i(\xb,\xi,t) =  \int_{-1}
     ^{1}\, d\beta \int_{-1+|\beta|}
     ^{1-|\beta|}\, d\alpha \delta(\beta+ \xi \, \alpha - \xb)
\, f_i(\beta,\alpha,t).
\end{equation}
The GPDs are related with PDFs through the double distribution
function
\begin{equation}\label{ddf}
f_i(\beta,\alpha,t)= h_i(\beta,t)\,
                   \frac{\Gamma(2n_i+2)}{2^{2n_i+1}\,\Gamma^2(n_i+1)}
                   \,\frac{[(1-|\beta|)^2-\alpha^2]^{n_i}}
                          {(1-|\beta|)^{2n_i+1}}\,.
                          \end{equation}
The powers $n_i=1,2$ (i= gluon, sea, valence contributions) and
the functions $h_i(\beta,t)$ are proportional to parton
distributions \cite{gk06}.

To calculate GPDs, we use the CTEQ6 fits of  PDFs for gluon,
valence quarks and sea \cite{CTEQ}. Note that the $u(d)$  sea and
strange sea are not flavor symmetric. In agrement with CTEQ6 PDFs
we suppose that $H^u_{sea} = H^d_{sea} = \kappa_s H^s_{sea}$, with
\begin{equation}\label{kapp}
\kappa_s=1+0.68/(1+0.52 \ln(Q^2/Q^2_0))
\end{equation}

The parton subprocess ${\cal H}^{V}$ contains a hard part which is
calculated perturbatively  and the $k_\perp$- dependent wave
function. It contains the leading and higher twist terms
describing the longitudinally and transversally polarized vector
mesons, respectively. The quark transverse momenta  are considered
in hard propagators decrease the $LL$ amplitude and the cross
section becomes in agrement with data. For the $TT$ amplitude
these terms regularize the end point singularities.

We consider the gluon, sea and quark GPDs contribution to the
amplitude. This permits us to analyse vector meson production from
low $x$  to moderate values of $x$ ($ \sim 0.2$) typical for
HERMES and COMPASS. The obtained results \cite{gk06} are in
reasonable agreement with experiments at HERA\cite{h1,zeus},
HERMES \cite{hermes}, COMPASS \cite{compass} energies for
electroproduced $\rho$ and $\phi$ mesons.
\begin{figure}[h!]
\begin{center}
\begin{tabular}{cc}
\includegraphics[width=7.7cm,height=5.9cm]{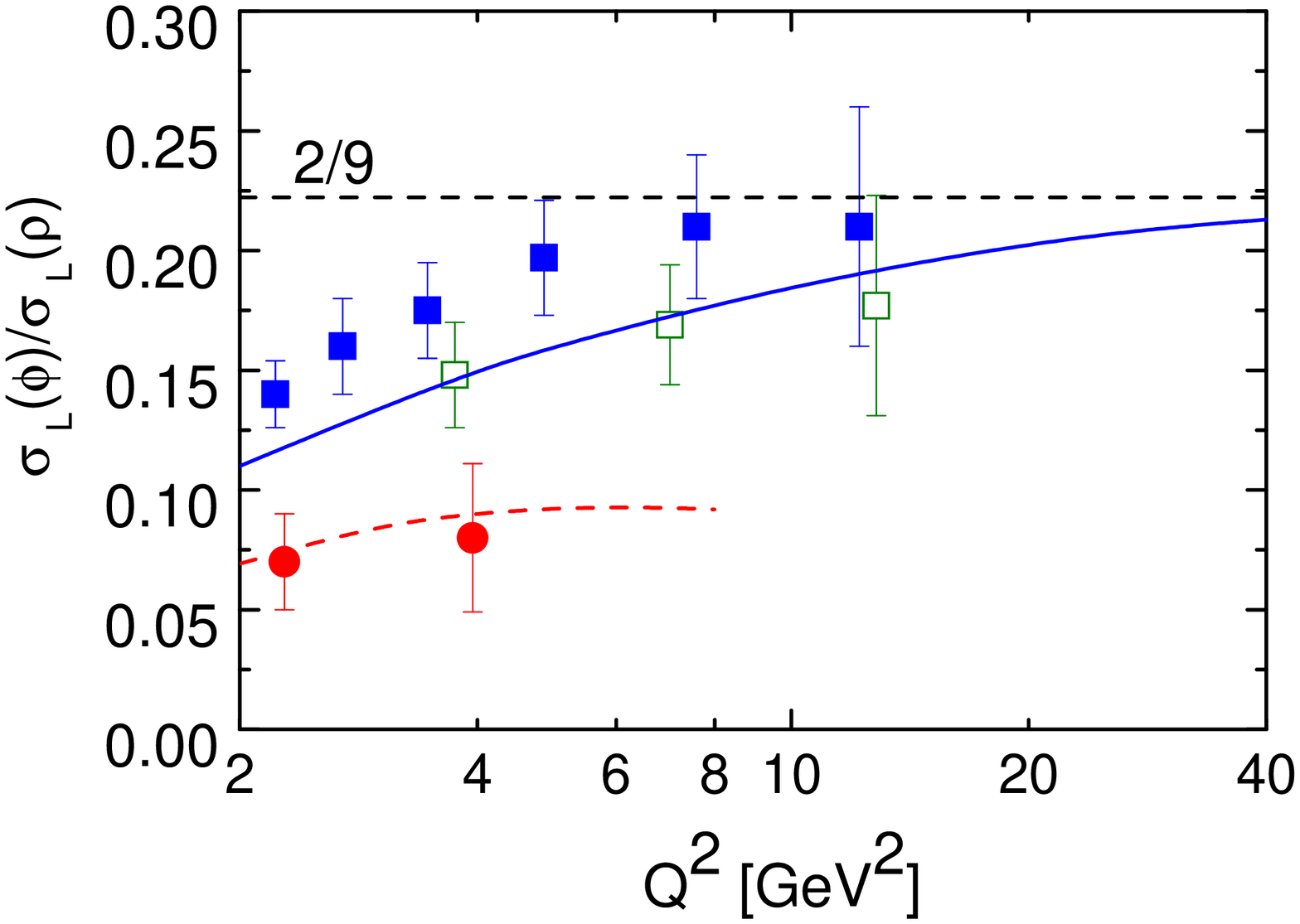}&
\includegraphics[width=7.7cm,height=5.8cm]{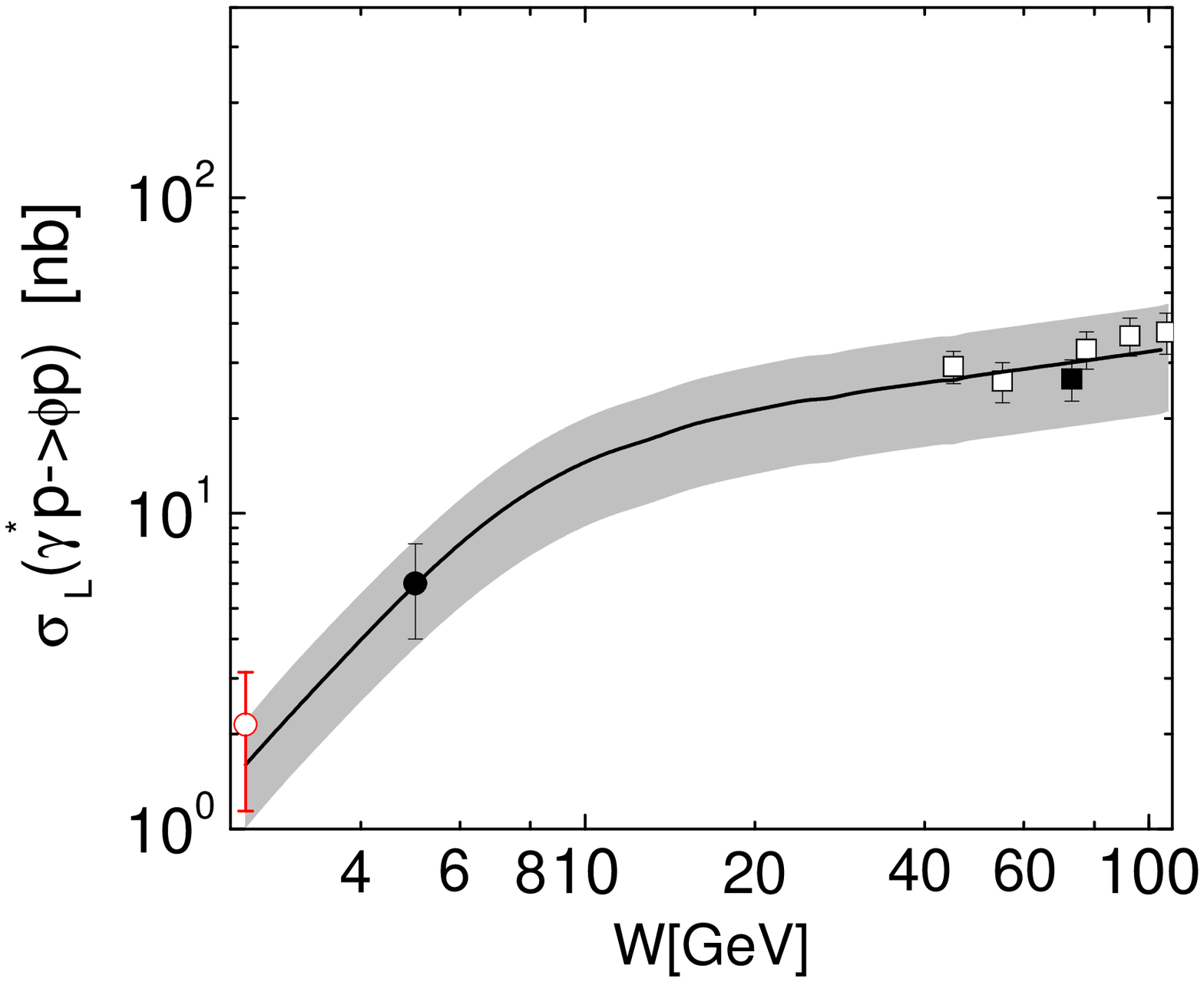}\\
\textbf{(a)} & \textbf{(b)}
\end{tabular}
\end{center}
  \caption{%
    \textbf{(a)} The  ratio of cross sections $\sigma_\phi/\sigma_\rho$ at
    HERA energies- full line and HERMES- dashed line. Data are from H1  -solid, ZEUS
 -open squares, HERMES solid circles.
    \textbf{(b)} The longitudinal cross section for $\phi$ at
$Q^2=3.8\,\mbox{GeV}^2$. Data: HERMES, ZEUS, H1, open circle- CLAS
data point.}
\end{figure}

In Fig~1a, we show the strong violation of the
$\sigma_\phi/\sigma_\rho$ ratio from 2/9 value at HERA energies
and low $Q^2$, which is caused by the flavor symmetry breaking
(\ref{kapp}) between $\bar u$ and $\bar s$. The valence quark
contribution to $\sigma_\rho$ decreases this ratio at HERMES
energies. It was found that the valence quarks substantially
contribute only at HERMES energies. At lower energies this
contribution  becomes small and the cross section decreases with
energy. This is in contradiction with CLAS results which innerve
essential increasing of $\sigma_\rho$ for $W<5 \mbox{GeV}$. On the
other hand, we found good description of $\phi$ production at CLAS
 \cite{clas} Fig~1b. This means that we have problem only with the
valence quark contribution at low energies.
\begin{figure}[h!]
\begin{center}
\begin{tabular}{cc}
\includegraphics[width=7.7cm,height=6cm]{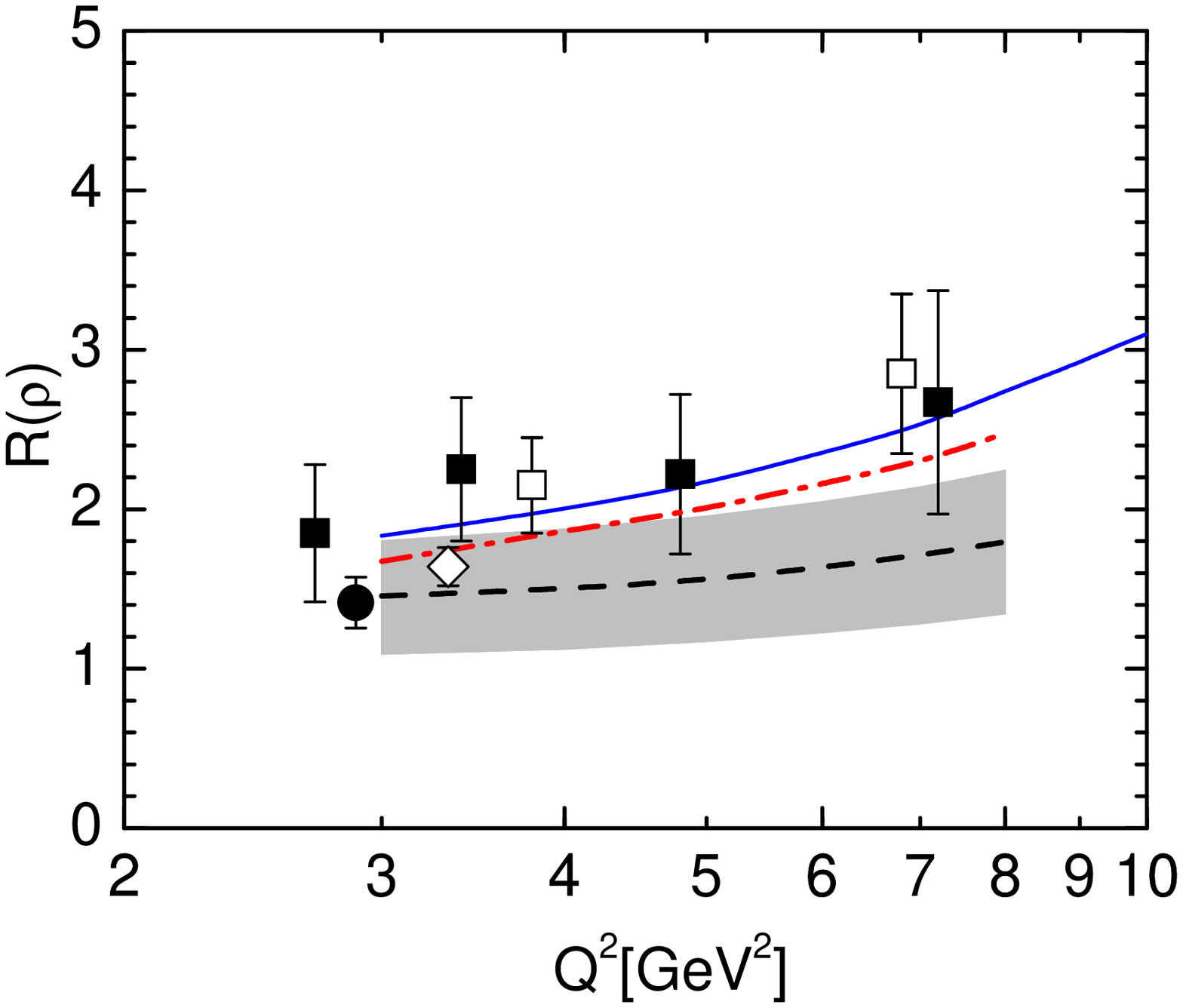}&
\includegraphics[width=7.7cm,height=6cm]{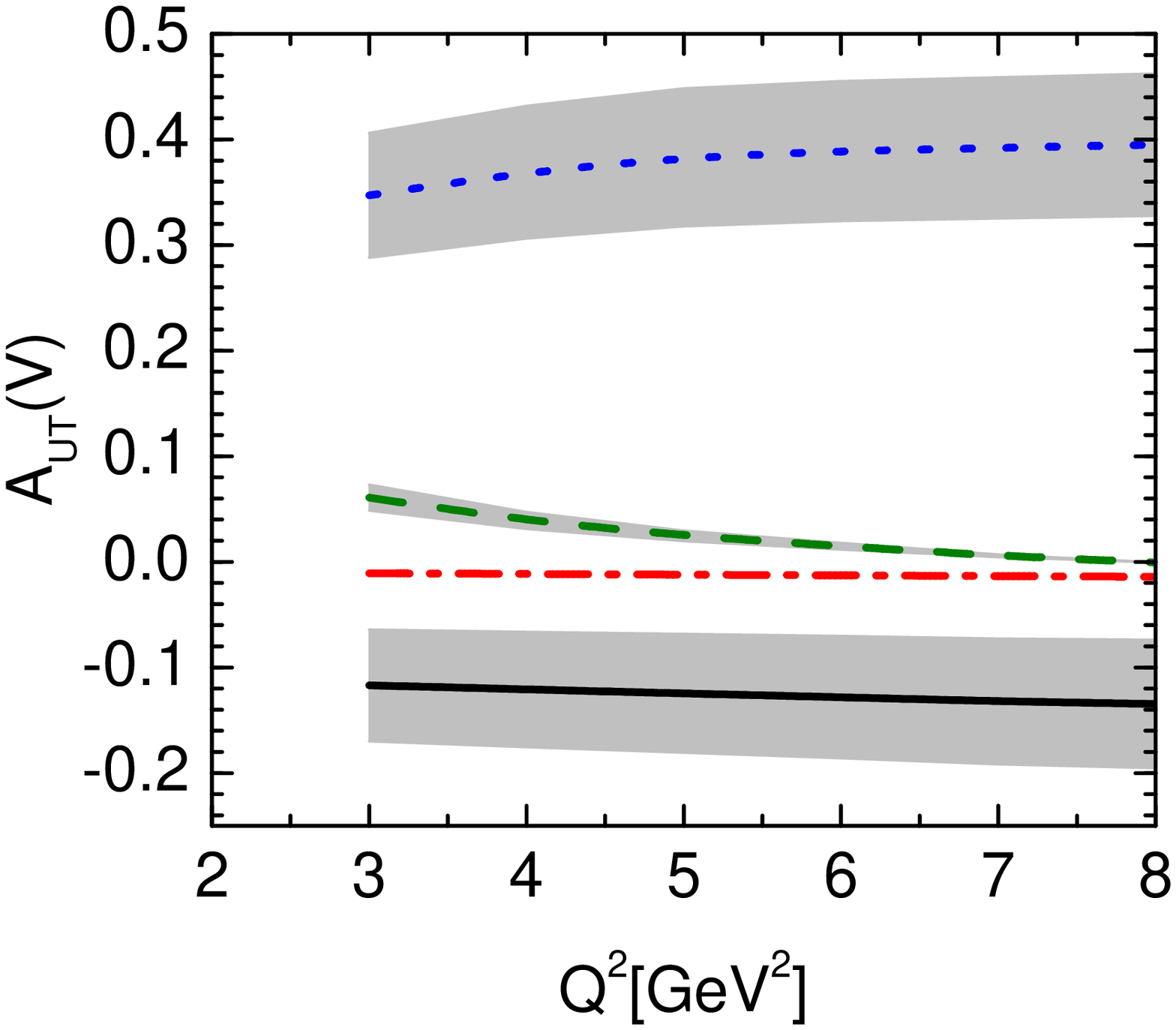}\\
\textbf{(a)} & \textbf{(b)}
\end{tabular}
\end{center}
  \caption{%
    \textbf{(a)} The ratio of longitudinal and transverse cross sections
 for $\rho$  production at low  $Q^2$ Full line- HERA ,
 dashed-dotted -COMPASS and dashed- HERMES.
    \textbf{(b)}  Predicted $A_{UT}$ asymmetry at COMPASS for various mesons.
    Dotted-dashed line $\rho^0$;
    full line $\omega$; dotted line $\rho^+$ and dashed line $K^{* 0}$.
  }
\end{figure}

\begin{figure}[h!]
\begin{center}
\begin{tabular}{cc}
\includegraphics[width=7.7cm,height=6cm]{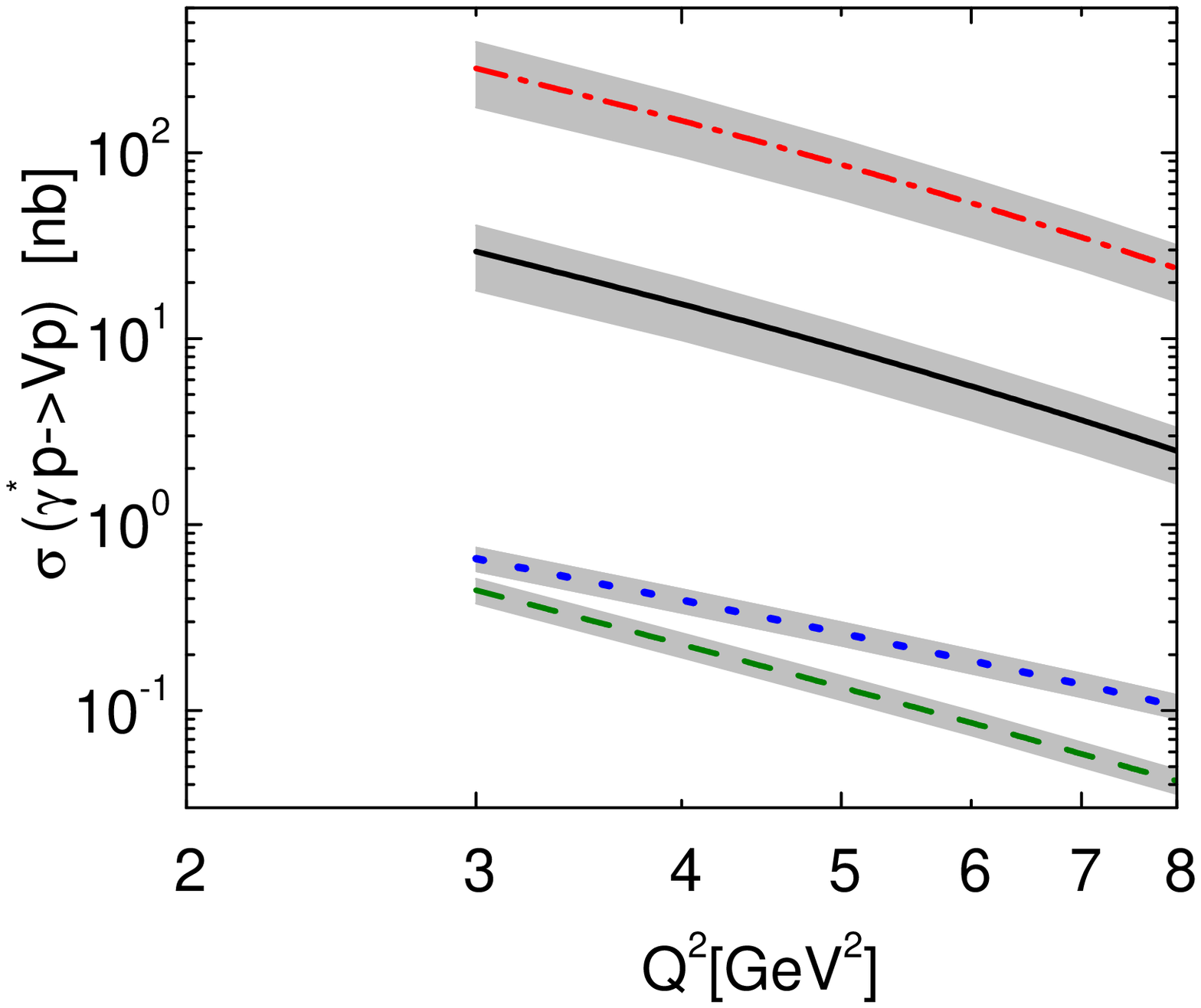}&
\includegraphics[width=7.7cm,height=6cm]{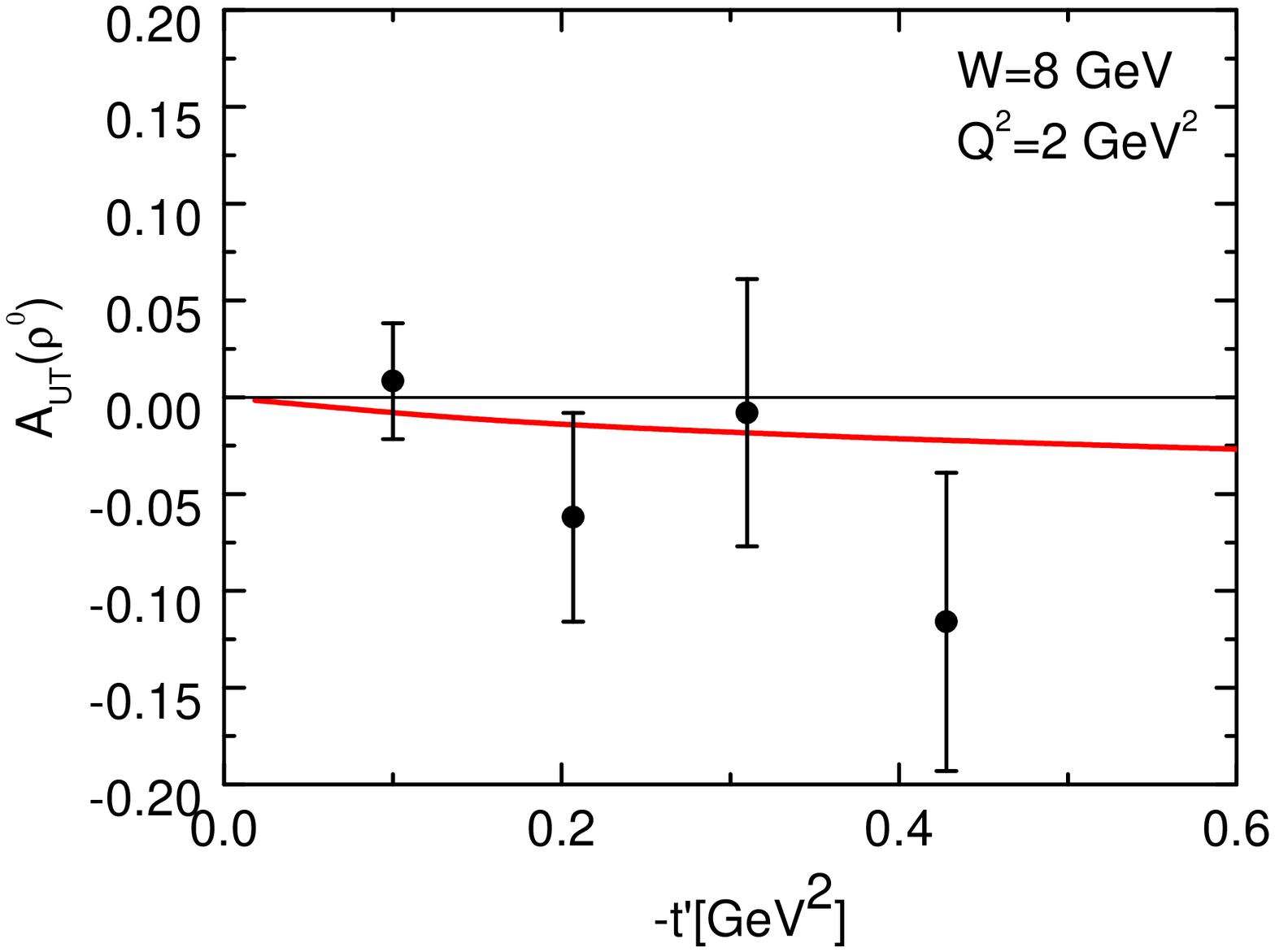}\\
\textbf{(a)} & \textbf{(b)}
\end{tabular}
\end{center}
  \caption{%
    \textbf{(a)} The integrated cross sections of vector meson
production at $W=10 \mbox{GeV}$. Lines are the same as in Fig~2b.
    \textbf{(b)}  Predictions for $A_{UT}$ asymmetry  $W=8 \mbox{GeV}$.
Preliminary COMPASS data at  this energy are shown \cite{sandacz}.
  }
\end{figure}
The results for the $R=\sigma_L/\sigma_T$ ratio are shown in
Fig~2a for HERA, COMPASS and  HERMES energies. We found that our
model describe fine $W$ and $Q^2$ dependencies of $R$.

The analysis of the target $A_{UT}$ asymmetry for
electroproduction of various vector mesons was carried out  in our
approach too\cite{gk08t}. This asymmetry is sensitive to an
interference between  $H$ and $E$ GPDs. We  constructed the GPD
$E$ from double distributions and constrain it by the Pauli form
factors of the nucleon, positivity bounds and sum rules. The GPDs
$H$ were taken from our  analysis of the electroproduction cross
section. Predictions for the $A_{UT}$ asymmetry at $W=10
\mbox{GeV}$ are given for $\omega$, $\phi$, $\rho^+$, $K^{*0}$
mesons \cite{gk08t} in Fig~2b. It can be seen that we predicted
not small negative asymmetry for $\omega$ and large positive
asymmetry for $\rho^+$ production. In these reactions the valence
$u$ and $d$ quark GPDs contribute to the production amplitude in
combination $\sim E^u-E^d$  and do not compensate each other
($E^u$ and $E^d$ GPDs has different signs). The opposite case is
for the $\rho^0$ production where one have the $\sim E^u+E^d$
contribution to the amplitude and valence quarks compensate each
other essentially. As a result $A_{UT}$ asymmetry for $\rho^0$ is
predicted to be quite small. Unfortunately, it is much more
difficult to analyse experimentally $A_{UT}$ asymmetry for
$\omega$ and $\rho^+$ production with respect to $\rho^0$ because
the cross section for the first reactions is much smaller compared
to $\rho^0$, Fig.~3a. Our prediction for $A_{UT}$ asymmetry of
$\rho^0$ production at COMPASS reproduces well the preliminary
experimental data Fig.~3b.

Thus, we can conclude that the vector meson electroproduction at
small $x$ is a good tool to probe the   GPDs. In different energy
ranges, information about quark and gluon GPDs can be extracted
from the cross section and spin observables of the vector meson
electroproduction.

 This work is supported  in part by the Russian Foundation for
Basic Research, Grant 09-02-01149  and by the Heisenberg-Landau
program.


\begin{thebibliography}{99} %% for less than 10 references use just {9}
\bibitem{fact} X.\ Ji, Phys. Rev. {\bf D55} (1997), 7114;\\
A.V.\ Radyushkin,  Phys. Lett. {\bf B380} (1996)  417;\\ J.C.\
Collins {\it et al.}, Phys.\ Rev. {\bf D56} (1997) 2982.
\bibitem{mp} L.\ Mankiewicz,  G.\ Piller,
   Phys.\ Rev. {\bf D61} (2000) 074013;\\
I.V.\ Anikin,  O.V.\ Teryaev, Phys.\ Lett. {\bf B554} (2003) 51.
\bibitem{gk06} S.V. Goloskokov, P. Kroll,
  Euro. Phys. J. {\bf C50}  (2007) 829; ibid {\bf C53}  (2008)
  367.
\bibitem{mus99} I.~V.~Musatov,  A.~V.~Radyushkin,
  Phys.\ Rev. {\bf D61} (2000) 074027.
\bibitem{CTEQ} J.~Pumplin, D.~R.~Stump, J.~Huston, H.~L.~Lai,
  P.~Nadolsky,  W.~K.~Tung, JHEP {\bf 0207} (2002) 012.

 \bibitem{h1} C.~Adloff  et al.  [H1 Collab.],
                        Eur.\ Phys.\ J. {\bf C13} (2000) 371;\\
   S.Aid et al. [H1 Collab.], Nucl. Phys. {\bf
                        B468} (1996)  3.
\bibitem{zeus} J.~Breitweg et al.  [ZEUS Collab.],
  Eur.\ Phys.\ J. {\bf C6}  (1999)  603;\\
 S. Chekanov et al.  [ZEUS Collab.], Nucl. Phys. {\bf B718}
 (2005) 3;\\
   S.~Chekanov et al.  [ZEUS Collab.],  PMC Phys. {\bf A1} (2007) 6.
\bibitem{hermes} A.~Airapetian  et al.   [HERMES Collab.],
     Eur.\ Phys.\ J.\ {\bf C17} (2000) 389;\\
      A. Borissov, [HERMES Collab.], "Proc. of Diffraction 06", {\bf
      PoS} (DIFF2006), 014.
\bibitem{compass} D.\ Neyret [COMPASS Collab.],  "Proc. of SPIN2004", Trieste,
Italy, 2004;\\
 V.~Y.~Alexakhin
et al. [COMPASS Collab.],  Eur. Phys. J. {\bf C52} (2007) 255.

\bibitem{clas}  J. P. Santoro et al.  [CLAS Collab.], Phys. Rev. {\bf C78} (2008)
025210.
\bibitem{sandacz} A.~Sandacz [COMPASS Collab.], this proceedings.
\bibitem{gk08t} S.V. Goloskokov, P. Kroll, Eur. Phys. J.
{\bf C59} (2009) 809.
\end{thebibliography}
\end{document}